\documentstyle[12pt]{article}
\begin{document}
\title{From GM law \\to \\A powerful mean field  scheme}
\renewcommand{\thefootnote}{\fnsymbol{footnote}}
\author{Serge Galam \\  (E-mail: galam@ccr.jussieu.fr)\\ $\,$\\
Laboratoire des Milieux D\'{e}sordonn\'{e}s 
et H\'{e}t\'{e}rog\`{e}nes\thanks{Laboratoire associ\'{e} au 
CNRS (UMR 7603)},\\
Tour 13 - Case 86, 4 place Jussieu, \\ 75252 Paris Cedex 05, France\\}
\date {J. App. Phys., in press, 2000}
\maketitle

\begin{abstract}

A new and powerful mean field scheme is presented. It maps
to a one-dimensional finite closed chain in an external field.
The chain size accounts for lattice topologies.
Moreover lattice connectivity is rescaled according
to the GM law recently obtained in percolation theory.  The associated
self-consistent mean-field equation of state yields critical 
temperatures which are within a few percent of exact estimates. Results are obtained 
for a large variety of lattices and dimensions.
The Ising lower critical dimension for the onset of phase transitions is $d_l=1+\frac{2}{q}$.
For the Ising hypercube it becomes the Golden number $d_l=\frac{1+\sqrt 5}{2}$.
The scheme  recovers the exact result of no long range order 
for non-zero temperature Ising
triangular antiferromagnets.

\end{abstract}
\newpage

\section{Introduction}

For many years basic mean-field theory has been applied to a huge variety of problems.
It is a very simple way 
to tackle collective phenomena [1]. Most of the time it yieds a correct qualitative description.
However, quantitatively  the results are very poor. In particular, all aspects of
the critical behavior are .grossly misrepresented [2]. Moreover some results are even wrong like
for instance the existence of long range order for the Ising system at both one dimension for ferromagnets
and at two dimensions for the triangular antiferromagnets.

The Bethe scheme [3] was then introduced to extend the Weiss one spin approach [4] to a 
cluster of fluctuating spins. For decades it has been looked upon as a solid imrovement. 
It yields  no long range order at one dimension and even becomes exact for the hypercube at
infinite dimensions [1]. However quantitative results are yet rather poor. 
Moreover it was demonstrated recently that the Bethe scheme violates systematically 
translational invariance [5]. As such it is forbidden by symmetry.

In this paper we  present a new and powerful mean field scheme. It embodies 
the Bethe idea of including a fluctuating spin cluster yet
preserving the overall lattice translational invariance. In addition, the connectivity between fluctuating 
clusters is rescaled according to the GM law introduced few years ago in percolation theory [6].

Associated critical temperatures are calculated for a large variety of
lattices and at several dimensions. Discrepancies with available exact estimates are only within 
few percent. 
The lower critical dimension for the onset of phase transitions is found to be $d_l=1+\frac{2}{q}$
for Ising systems. It turns to he Golden number $d_l=\frac{1+\sqrt 5}{2}$ for hypercubes ($q=2d$).
In the case of the triangular Ising antiferromagnet the exact result of no long range order 
is reproduced [7]

\section{Revisiting the Bethe aprroximation}

Mean-field theory is a one-site approach which first breaks the lattice symmetry by discriminating
bewteen fluctuating degrees of freedom and
averaged ones [1].
Two interpenetrated lattices are thus defined. Equating the thermal average of fluctuating degree of 
freedom to the already averaged ones restores the inital lattice symmetry. Simultaneously  
a self-consistent equation of state is obtained.

To implement a Bethe scheme [3] on a lattice 
3 distinct interpenetrated sublattices (A, B, C) must be introduced. First the fluctuating center (A), 
then the fluctuating nearest neigbhors (B) and last the mean field nearest neigbhors (nn) of the nn (B)
not including the center (A).
From the A-spin plus these B and C shells, a cell is constituted to pave the whole 
space and reproduce the full lattice topology.

Cluster center (A) has thus all its nn spins (B) which are fluctuating while
surface cluster spins (B)
have mean-field nn spins (C) and one nn fluctuating spin (A). Simultaneously mean-field spins (C) have 
all their 
nn which are fluctuating spins, making their environnement identical to the cluster center. 

On this basis
the Bethe requiremnt $<S_A>=<S_B>$ is not compatible with the equality $m_C=<S_B>$ which should 
also hold to ensure translational invariance. The Bethe topology is therefore forbidden 
by symmetry. It is not the case for one-site Weiss theory. For a detailled demonstration see [5]

Last but not least, it is worth noticing it is indeed this very symmetry problem 
which makes the Bethe approach exact on the Caley tree lattice.
This lattice does not exhibit translational invariance by construction. 
This symmetry breaking  was overlooked for several decades.

\section{A new powerful mean field schme}

From the discovery of a systematic Bethe induced symmetry breaking arises the question 
of the possibility to indeed
extend a mean field treatment to more than one site. 

Above analysis of the Bethe scheme emphazises the role of the cluster center in the irreversible 
breaking of the symmetry. It hints to avoid such a fluctuating center. One way to achieve
this constraint is to 
use compact closed linear loops within the lattice topology. For instance compact
4-spin squares and 3-spin triangles for respectively square and triangular lattices. 

Each one of these plaquettes is then
set respectively as A-species (fluctuating) and B-species (mean field) with
a staggered-like coverage pattern. A-plaquettes (B-plaquettes) have thus all their nn plaquettes
as B-plaquettes (A-plaquettes).

For a given plaquette, each spin has two nn spins of the same
species within the plaquette itself and $(q-2)$ nn spins of the other species belonging to nn
plaquettes. At this stage we have a series of
fluctuating one-dimensional closed 
chains in an external field $h$. The number of spins $N$ in each chain is determined from lattice topology.
It is $N=4$, $N=3$, $N=3$ and $N=6$ for respectively square, triangular, 
Kagom$\acute{e}$ and Honeycomb lattices. 

it is the interactions with nn mean-field spin plaquettes which produce the field $h$. We have $h=\delta Jm$ 
where
$\delta$ accounts for connectivety to B-sublattices, $J$ is the nn coupling constant and $m$ 
the averaged magnetization on the B-sublattice. The problem can now be solved exactly.
In particular, the chain site magnetization is [1],   
\begin{equation}
<S_i>=\beta \exp{2K}\{\frac{(1-\tanh(K)^N)}{(1+\tanh(K)^N)}\}h\ ,
\end{equation}
at order one in $h$. Here $i\in A$-plaquettes. $K\equiv \frac{J}{k_BT}$ where $k_B$ is
the Boltzman constant and $T$ the temperature.

Putting $<S_i>=m$ restores the initial lattice symmetry. It
is indeed possible since  
only two sublattices were involved which was not the case for the 3 sublattice Bethe scheme.
The self-consistent equation of state is,
\begin{equation}
m=\delta K \exp{2K}\{\frac{(1-\tanh(K)^N)}{(1+\tanh(K)^N)}\} m+... \ ,
\end{equation}
at order one in $m$ and using $h=\delta Jm$.
To solve Eq. (2) needs to determine the value of $\delta$. 

It is then worth  to evoke a recent work on percolation thresholds, the GM law [6]. It shows that 
relevant connectivity variables for site and bond dilution are repectively
$(d-1)(q-1)$ and $\frac{(d-1)(q-1)}{d}$. In other words, for site percolation, 
the number of possible directions $(q-1)$ from a given site, has to be multiplied by $(d-1)$.
For bond percolation this effective number of site directions has to be divided by dimension $d$.
Using these variables, the GM law was found to yield all percolation thresholds
for all Bravais lattices at all dimensions [6].

This percolation finding suggests to consider here a rescaled connectivity between closed loops
instead of $\delta =q-2$. Using above counting, we first start with $q$ instead of $(q-1)$
since now dealing with pair exchange interactions and not percolation. Second we
renormalize $q$ by $(d-1)$ giving $q(d-1)$. However the $2$ neighboring
sites which are treated exactly within the closed loop have to be substracted from the effective
number of sites which gives 
$q(d-1)-2$. Moreover, interactions being related to bonds, we divide this number by $d$ 
as for bond percolation. These considerations  lead to a connectivity, 
\begin{equation}
\delta =\frac{q(d-1)-2}{d} \ . 
\end{equation}

\section{ Results}
We can now check the validity of our simple symmetry preserving model with respect to critical 
temperatures.
From Eq. (2) we get, 
\begin{equation}
\delta K_c^G \exp{2K_c^G}\{\frac{(1-\tanh(K_c^G)^N)}{(1+\tanh(K_c^G)^N)}\}=1\ .
\end{equation}
The trivial connectivity counting $\delta=q-2$ already improves Weiss model. For instance 
$K_c^G=0.29$ in the square case and $T_c=0$ at $d=1$. We now proceed using Eq. (3) 
for connectivty.

For the square case ($q=4,\ N=4$), $\delta=1$ which gives
$K_c^G=0.4399$. Exact result is $K_c^e=0.4407$. In the case of triangular 
lattice ($q=6,\ N=3$),  $K_c^G=0.2919$ with $\delta=2$ while the exact
estimate is $K_c^e=0.2746$. For Kagom$\acute{e}$ ($q=4,\ N=3$) $\delta=1$ yielding
$K_c^G=0.4649$ for an exact estimate of $K_c^e=0.4666$. And $K_c^G=0.6160$ for the honeycomb 
lattice
($q=3,\ N=6$) where $\delta=\frac{1}{2}$ for an exact estimate of
$K_c^e=0.6585$ (see Table I).

Going to $d=3$ imposes to restrict the plaquette size to $N=4$ since a 
one-dimensional loop cannot embody a three-dimensional topology. However there exits a $d-$dependence
through Eq. (3). We get $\delta=\frac{10}{3}$, $\delta=\frac{14}{3}$ and 
$\delta=\frac{22}{3}$ for respectively cubic, fcc and bcc lattices.
Corresponding critical temperatures are given by 
$K_c^G= 0.2012,\  0.1568,\  0.1096$ respectively for exact estimates of 
$0.2217,\  0.1575,$ and $0.1021$ (see Table I).

Critical temperature estimates [8, 9] are available for the hypercube at $d=5,\  6,\ 7$. These are
$K_c^e= 0.1139,\ 0.0923,\ 0.0777$ respectively.

To get the $d \rightarrow \infty$ asymptotic limit of our model
we take both $q \rightarrow \infty$ and  $J\rightarrow 0$ 
under the constraint $qJ=cst$.
From Eq. (3) connectivity limit is
$\delta \rightarrow q(1-\frac{1}{d})$ which gives always,
\begin{equation}
\delta \rightarrow q \,
\end{equation}
at leading order. Indeed $q$ diverges always quicker than 
$\frac{q}{d}$.even for $fcc$-lattices where $q=2d(d-1)$.  
In turn  Eq. (4) becomes, 
\begin{equation}
K_c^G  =\frac{1}{q}\ ,
\end{equation}
which is the mean-field 
result [1] as expected in the $d\rightarrow \infty$ limit.

To evaluate the sensibility on the loop size, it is fruitful to
expand Eq. (4) in powers of $K$. It gives
\begin{equation}
K_c^G (1+2K_c^G+...+\frac{(2K_c^G)^N}{N!})(1-(K_c^G)^N+...) (1-(K_c^G)^N+...)= \frac{1}{\delta }\ .
\end{equation}
Since $N\geq 3$, a simple analytic expression is obtained only at order one,
\begin{equation}
K_c^G  =\frac{1}{\delta}\ .
\end{equation}
At two dimensions Eq. (8) gives $K_c^G=1,\ \frac{1}{2},\ 1,\ 2$ for respectively the square, 
triangular,  Kagom$\acute{e}$ and for honeycomb lattices.
These results are rather poor and shows the importance of the finite value of $N$ which 
embodies part of the lattice topology.

\section{Conclusion}

We have presented a very simple self-consistent model which yields 
rather good values for critical temperatures within a few percent of exact results. 
Besides a rescaled lattice connectivity, the finite
length of the loops is also taken into account. This new scheme represents a substantial
improvement over existing mean-field cluster approximations. 

We can also determine from our model a lower critical dimension for phase 
transitions. 
It comes from the condition $h=0$ for which we have a one-dimensional finite system.
Such a system has  no long 
range order at $T \neq 0$. Phase transitions are thus obtained only in the range $h \neq 0$
which gives $q(d-1)> 2$ leading to,
\begin{equation}
d_l=1+\frac{2}{q}\ .
\end{equation}
For the Ising hypercube ($q=2d$) it becomes the Golden number $d_l=\frac{1+\sqrt 5}{2}$,
which excludes the $d=1$ case and contains $d=2$ as it should be.
Last but not least, applying our scheme to the Ising
triangular antiferromagnet[7] we do
revover the exact result of no long range order 
at non-zero temperatures; contrary to usual mean field approaches.

\newpage
{\LARGE References}\\ \\

$\ $\\
1. {\sf R. K. Pathria}, Statistical Mechanics, Pergamon Press (1972)  \\
2. {\sf Sh-k Ma},  Modern Theory of Critical Phenomena, The Benjamin Inc.: Reading MA (1976) \\
3. {\sf H. A. Bethe}, Proc. Roy. Soc. London A\underline {150}, 552 (1935)  \\
4. {\sf P. Weiss}, J. Phys. Radium, Paris \underline {6}, 667 (1907)  \\
5. {\sf  S. Galam}, Phys. Rev. B\underline {54}, 1599 (1996)  \\
6. {\sf  S. Galam and A. Mauger}, Phys. Rev. B\underline {53}, 2171 (1996)  \\
7. {\sf  S. Galam and P. V. Koseleff}, to be published (2000)  \\
8. {\sf M. E. Fisher}, Repts. Prog. Phys. V\underline {XXX} (II), 671 (1967)  \\
9. {\sf  J. Adler}, in ``Recent developments in computer simulation studies 
in Condensed matter physics", VIII, edited by D. P. Landau, Springer (1995) \\

\newpage
\begin{table}

\label{tbl}
\begin{center}
\begin{tabular}{|l|l|r|r|r|r|}
\hline  
Dimension &Lattice &$\,q$&$\delta$&${K_c^e}$&${K_c^G}$ \\ [5pt] 
\hline 
$d=2$ &Square & 4 &1&  0.4407& 0.4399  \\ [5pt]
$\,$ &Honeycomb & 3&$\frac{1}{2}$&  0.6585& 0.6160\\ [5pt]
$\,$ &Triangular & 6&2&  0.2746& 0.2837 \\ [5pt]
\hline
\hline
$d=2$ &Kagom$\acute{e}$*&  4&1& 0.4666 & 0.4649 \\ [5pt]
\hline
$d=3$ &Diamond & 4& 2&0.3698& 0.2857\\ [5pt]
$\,$ &sc& 6&$\frac{10}{3}$&  0.2216& 0.2012\\ [5pt]
$\,$ &bcc& 8&$\frac{14}{3}$& 0.1574& 0.1568\\ [5pt]
$\,$ &fcc& 12&$\frac{22}{3}$ & 0.1021& 0.1096\\ [5pt]
\hline
$d=4$ &sc& 8&$\frac{22}{4}$& 0.1497& 0.1380\\ [5pt]
$\,$ &fcc& 24 &$\frac{23}{2}$& $\ $& 0.0749\\ [5pt]
\hline
$d=5$ &sc& 10&$\frac{38}{5}$& 0.1139& 0.1064\\ [5pt]
$\,$ &fcc& 40&$\frac{158}{5}$ & $\ $& 0.0298\\ [5pt]
\hline
$d=6$ &sc& 12&$\frac{29}{3}$& 0.0923& 0.0869\\ [5pt]
\hline
$d=7$ &sc& 14&$\frac{82}{7}$& 0.0777& 0.0737\\ [5pt]
\hline
\end{tabular}
\end{center}
\caption{\sf $K_c^G$ from this work compared to ``exact estimates" 
$K_c^e$ taken from [8, 9].} 
\end{table}

\end{document}